\documentclass[twocolumn,aps,amsmath]{revtex4}
\usepackage[dvips]{graphics}
\newcommand{\be}{\begin{equation}}
\newcommand{\ee}{\end{equation}}

\newcommand{\bea}{\begin{eqnarray}}
\newcommand{\eea}{\end{eqnarray}}
\newcommand{\bd}{\begin{displaymath}}
\newcommand{\ed}{\end{displaymath}}
\newcommand{\bi}{\begin{itemize}}
\newcommand{\ei}{\end{itemize}}
\newcommand{\bc}{\begin{center}}
\newcommand{\ec}{\end{center}}
\newcommand{\bfl}{\begin{flushleft}}
\newcommand{\efl}{\end{flushleft}}
\newcommand{\bfr}{\begin{flushright}}
\newcommand{\efr}{\end{flushright}}

\def\6{\partial}

\def\={\!\!\!&=&\!\!\!}
\def\+{\!\!\!&&\!\!\!+~}
\def\-{\!\!\!&&\!\!\!-~}

\begin{document}
\title{Metallic Glass in Two Dimensional Disordered Bose System; A Renormalization
Group Approach.}
\author{M. Crisan}
\affiliation{Department of Theoretical Physics, University of
Cluj, 3400 Cluj-Napoca, Romania}
\author{I. Grosu}
\affiliation{Department of Theoretical Physics, University of
Cluj, 3400 Cluj-Napoca, Romania},
\author{I.Tifrea}
\affiliation{Department of Theoretical Physics, University of
Cluj, 3400 Cluj-Napoca, Romania}
\affiliation{Department of Physics and Astronomy, The University
of Iowa, Iowa City, IA 52242, USA}
%\maketitle

\begin{abstract}
We consider the two dimensional disordered Bose gas which present
a metallic state at low temperatures. A simple model of an
interacting Bose system in a random field is propose to consider
the interaction effect on the transition in the metallic state.
\end{abstract}

\maketitle

\section{Introduction}

 The recent experimental data obtained on the two-dimensional (2d)
 systems \cite{exp1, exp2, exp3} showed the possibility of
 the occurrence for the metallic phase in the
 insulator-superconductor transition. The model proposed by Das and
 Doniach  \cite{das} is a uniform phase lacking both phase and
 charge order. This phase presents translationally and rotationally
 invariance. It has to appear at finite temperature, because at $T=0$
 there is no phase transition which is lacking both charge and
 phase order \cite{fat}. The occurrence of the metallic phase
 glass in a disordered two dimensional ($2d$) bosonic system has
 been predicted by Wagenblastl et al.  \cite{wag} in the
 insulator-superconductor transition, but only at the separatrix
 line. The important result has been obtained by Dalidovich and
 Phillips  \cite{dal1}, who shoved that this state can appear
 in  disordered two dimensional  $(2d)$ Bose systems. This model is based on the
 quantum version of the spin glass state which can be described
 using the replica trick method. The initial free energy has the
 universality class $z=1$ ($z$ is the dynamical critical exponent)
 with a local breaking spin rotation
 invariance. The effective order parameter , is given in terms of
  the phase $\varphi(r,\tau)$ ($\tau$ is the imaginary time)  as ,
   $Q(\tau)=<\exp(i{\varphi (r_{ij}},0)\exp(i{\varphi(r_{ij},\tau))})>$
 and relaxes exponentially fast to its equilibrium value. The quantum
 decay of $Q(\tau)$ is like $\tau^{-2}$ and this gives rise to
 excitations which scales as $|\omega|$ and changes the universality
 class  to $z=2$ \cite{pi1}. Such a dissipative model has been treated in
 \cite{dal1} in the Gaussian approximation and the conductivity $\sigma (T=0,\omega=0)$
 has been obtained as finite, which proves that the state is
 metallic.

  In this paper we present an equivalent model for the $2d$
  disordered Bose system which can be treated with the
  Renormalization Group (RG) method.  The model
  was  used to describe the spin glass
  state \cite {aa,me} where disorder was introduced by a random
  field which can be connected to the effective order parameter
  of the spin glas state. The quantum version of this method was given in  \cite {bu1,bu2}
  where was applied for the study of the critical behavior of the
  interacting Bose  system. However, we mention that the theory of
  the critical behavior of the transport is not trivial because at
  finite temperature the limits $\sigma(T=0, \omega)$ and $\sigma(\omega=0, T)$
  are not equal \cite{da1,sa},  but we will be interested in the
  calculation of the $\sigma(0,0)$ near the critical region.
    The model of our disordered Bose system will be presented in
    Sec.II, where we will discuss the equivalence with the model
    from Ref. \cite{pi1}. In Sec.III we solve the flow equations of
    the RG.equations of this model. The strategy of our
    calculations is different to this from \cite{bu1,bu2}, because
    we are interested in the case of $d=2$ and $z=2$. In this case
    we can solve the differential equations exactly, studding the
    influence of the temperature on the transition. In fact this
    procedure give us the possibility to analyze the relevance of
    the parameters $T$ and the interaction $u_0$ near the fixed point
     $T=0$. We expect that
    the new result for conductivity to contain a correction given
    by the interactions between bosons which is in fact small and are in fact
    corrections to the Gaussian behavior.
     The conductivity near the
    critical region will be calculated following the method given
    in  \cite{dal1}, and applied recently \cite{ti} for the study of the
     study of the fluctuations of the conductivity in the d-wave superconductors near the critical disorder.
      The last section, Sec.IV, will be devoted to the calculation
      of the conductivity and the result will be compared with the
      simple Gaussian model.
    . The values given by the experimental
    are different from the standard result obtained
    for the simple model which gives $\sigma(0)=4e^2/h$ ( 2e is the pair charge and h is the Planck constant)
     the well known result
     of Fisher and Grinstein \cite{fi}. This difference is not yet
     clear, but the accuracy of the measurements , as well as, the
     interactions from the systems can generate this deviation.
    The present results confirms the exciting idea of the
    existence of a metallic state in the insulator-superconductor
    transition which is driven by disorder, which in fact is not
    favorable for the superconducting state. We expect that this
    intermediate metallic state to appear in the the d-wave
    superconductors containing  nonmagnetic impurities. The
    model from Ref.\cite{ti} contains also the influence of the
    temperature on the  metallic conductivity near the critical concentration which destroys
    d-waves supercoducting state.
\section{Model.}
 Before the presentation of the model of an interacting Bose system the in a random field
 we start with a short presentation of the model proposed  by
 Dalidovich and Phillips \cite{dal1,pi1},which contains the
 physics of the problem. This discussion is also relevant to make
 clear the validity and advantage of the model proposed in this
 section.

\subsection{Dalidovich-Phillips model}
The Landau action for this problem can be obtained using the
replicas to perform the average on the disordered state.The
quadratic and quartic terms describe, in the spin glass theory the
interaction between spins ,that appears by the average procedure
on the disordered state,  and can be decoupled by introducing the
auxiliary fields $Q_{\mu \nu}^{ab}( \bf k, \bf k', \tau,\tau')=
<S_{\mu}^{a} S_{ \nu}^{b}( \bf k',\tau)>$ and $\Psi_{\mu}^{a}(\bf
k,\tau )=<S_{\mu}^{a}>$, where the superscriptions $a$ and $b$
represent the replica indices. A finite value of $\Psi_{\mu}^{a}$
 is equivalent with the phase ordering in the charge of $2e$
condensate, and $<\Psi_{\mu}^{a}>=0$ describe the disordered
phase. As we mentioned in introduction for the quantum spin glass
the diagonal elements of  $Q(\bf k,\bf k',\tau,\tau')$, in the
limit $|\tau-\tau'|\rightarrow \infty $, are the effective order
parameter.  The behavior of this parameter leads to the change of
the critical dinamic exponent $z=1$ to $z=2$. The free energy
given in \cite{dal1} contain a Gausssian and a quartic term as in
the standard $\Phi^{4}$-theory, and the contribution which couples
the charge and the glassy degrees of freedom, and is considered by
the authors as dominant in the bicritical point. However,using
such a form for action the application of the RG method seems to
be very difficult, and even the authors considered only the
Gaussian approximation.

 We will show that this state,studied in \cite{dal1} can be
 modelled  by an action which describes the effect of a random
 field on the interacting Bose system, using the RG in the quantum
 limit. In the following we will present the model which gives a
 similar result with the Gaussian model containing the corrections
 given by the interaction between bosons. The effect of disorder is
 contained only in the gaussian contribution , and we will show
 that in the lowest order the gaussian term is identical with the
 expression from \cite{dal1}.

\subsection{Random field model}
We consider a $d$-dimensional Bose model in a random field $h(\bf
x )$ described by the Hamiltonian:
\begin{widetext}
\begin{equation}
\textit{H}=\int d^d x \Phi^{\dag}(\bf x)[-\nabla^2+r_o]|\Phi(\bf
x)|^+ \frac{u_0}{4}\int d^d x |\Phi(\bf x)|^4 +\int d^ d x[h(\bf
x)\Phi^{\dag} (\bf x )+ h^{\star}(\bf x)\Phi(\bf x)].
\end{equation}
\end{widetext}

In this Hamiltonian ($\hbar^{2}=2m=1$), $\Phi(\bf x)$is the
bosonic field, $r_0$ the control parameter (for the standard case
it is the chemical potential ), and $u_0>0$ is the bare coupling
constant. The random field $h(\bf x)$ is a Gaussian random
variable with a Fourier transform $h(\bf k)$ which satisfies:
\begin{align}
<h(\bf k)>= <h^{\star}(\bf k)>=0, <h^{\star}(\bf k) h(\bf
k')>=\delta_{\bf k,\bf k'}q
\end{align}
where $<>$ indicates an average over the possible configurations
of the random field and will serve to introduce a an effective
Edwards-Anderson \cite {ea} spin-glass parameter denoted by
$q$.This Hamiltonian has been used to develop a functional theory
\cite{bu1, bu2} described by the action :
\begin{equation}
S[\Phi]=S^{(0)}[\Phi]+S^{(in)}[\Phi]
\end{equation}
where:
\begin{equation}
S^{(0)}[\Phi]=\sum^m _{a,b=1} \sum_k
[(r_0+k^2-i\omega_n)\delta_{ab}-(q/T)\delta\omega_{n0}]\Phi_a(k)\Phi_b(k)
\end{equation}
In this equation $k\equiv (\bf k,\omega_n) $ the indices a, b and
m are the replicas indices from the standard spin-glass theory
\cite{ea}, method used in \cite {bu1, bu2}  to calculate the free
energy of this disordered system. In the replica-trick theory the
calculations are performed taking $m\longrightarrow 0$.
 The interaction contribution has the form:
\begin{equation}
S^{(in)}[\Phi]=\frac{u_0}{4}\sum^m _{a=1}\sum_{k_1}...\sum_{k_4}
\Phi_a(k_1)...\Phi_a(k_4) \delta(k_1+...+k_4)
\end{equation}
The Gaussian propagator in the limit $m\longrightarrow 0$ has the
form \cite{pi1, bu1}:
\begin {equation}
G_{a,b}(k)=G_{0}(k)\delta_{a,b}+\beta
G^2_{0}(k)q\delta_{\omega_n,0}
\end {equation}
where $\beta=1/T$ and $G_0(\bf k,\omega_{n})=(r_0+k^2
+\eta\|\omega_{n}|)^{-1}$ .We mention that the damping term linear
in energy has been introduced in the model following the physical
considerations presented in \cite{da1, pi1}, and it has a
 importance, because it keeps the universality class  $z=2$ for
 the model.

In the next section we will use the results from \cite{bu1}to
write the flow equations for the model in the limit
$m\longrightarrow 0$. We also considered the replica symmetric
case, as in Ref.\cite{da1}, but we do not have a term which coupes
the bosonic field $\Phi$ and the order parameter $q$, which in
this model has been considered only spatial dependent. This may be
considered as a very poor approximation, because we used the an
free propagator with the Ohmic dissipative term, which implies a
time dependence of the effective parameter $q$. Anyway, even this
simple model, which is tractable by RG- theory,  contains more
than the Gaussian model, and can give us an idea of the simple
approximation performed in \cite {da1}.

\section{Renormalization group equations}

 The flow equations for the RG. can be obtained using the
recursion relations \cite{bu1} (where we take $b=e^l$) as:

\begin{equation}
\frac{dT}{dl}=2T.
 \end{equation}

\begin{equation}
\frac{dq}{dl}=4q
\end{equation}
\begin{equation}
\frac{dr}{dl}=2r+uF_1+uqF_2
\end{equation}
\begin{equation}
\frac{du}{dl}=[4-(d+z)]u-\frac{u^2}{4}[8F_3+2F_4]+5u^2q
\end{equation}
 In these equations we used for the number of components of the
 bosonic field $n=2$,and we will take $d=2$ and $z=2$. The
 functions $F_i,$($i=2...5$) can be calculated \cite{bu1}, (in this case there is a difference because of the
  damping $\eta$) and only the function
  $F_1(T(l)=K_2[\exp(1/T(l))-1]^{-1}$ ($K_2=1/(2\pi)$) is important because contains the
  relevant parameter $T(l)$. The Eqs.(9-10) will be solved in the
  low temperature limit and when we can take: $8F_3+2F_4 \simeq K_2$
 , and$F_5\simeq K_2$.

The solutions of the Eqs.(7-8) are $T(l)=Te^{2l}$ and
$q(l)=qe^{4l}$. Using these results we solved exactly the Eq.(9)
and the solution is:
\begin{equation}
u(l)=\frac{4}{K_2}\frac{1}{l+l_0+(e^{4l}-1)}.
\end{equation}
where$l_0=4/K_2u_0$.

 The general solution of the Eq.(9) has the form :
 \begin{equation}
r(l)=e^{2l}[r_0+I_1(T)+4q\int_0^l
dl'\frac{e^{2l'}}{l'+l_0+5q(e^{4l}-1)}],
 \end{equation}
where $I_1(T)$ is given by the expression:
\begin{equation}
I_1(T)=4\int_0^l dl'\frac{\exp{(-2l')}}{\exp(1/T(l')-1)}
\end{equation}
 We will consider the case of low temperatures when $I_1$ can be
 neglected,and $q$ is small, so we can approximate Eq.(12) by :
\begin{equation}
r(l)=e^{2l}\left[r_0+4q\int_0^l dl'\frac{e^{l'}}{l'+l_0}\right]
\end{equation}
This integral can be given by the function $Ei(x)$ defined by
\begin{equation}
Ei(xy)= e^{xy}\int_0^{1}dt\frac{ t^{y-1}}{x+\ln t}
\end{equation}
For $y=1$ and big argument $Ei(x)\simeq e^{x}/x$ and we get for
Eq.(14) the expression:
\begin{equation}
r(l)\simeq e^{2l}\left[r_0 +
2q\left(\frac{e^{2l}}{l+l_0}\right)-\frac{1}{l_0}\right]
\end{equation}

  Next important step is to calculate the stop scaling parameter
  $l^{\star}$ from the condition $r(l^{\star})=1$. This procedure
  applied first in \cite{da2} gives the possibility to study the
  influence of the temperature on the critical behavior of the
  transport properties in the critical region of the $z=2$
  insulator-superconductor transition. In our case the  Gaussian fixed point
  $T=0$ is stable, and we will expect a correction given by the
  interaction between fluctuations on this behavior. Indeed from
  Eq.(16) we obtain the important result:
  \begin{equation}
e^{4l^{\star}}\simeq (1+u_0q)^2
  \end{equation}

a result which will be used to calculate the conductivity of the
system considered, at $T=0$. The conductivity will be calculated
using the Kubo formula generalized for the  replicated action (1)
which contains the influence of disorder. In the next section we
will calculate $\sigma (i\omega_n)$ following the approximations
from \cite{da1} but using the renormalized value for $q(l)$ which
is a relevant parameter. This parameter is strongly dependent of
the variable $l$ and in this way we can fix it near the critical
point studding the competition between disorder and the
interaction between fluctuations in the occurrence of the metallic
phase.
 Before starting this calculation, which will follow the
 approximations from \cite {dal1}, we mention that even if our model
 is slightly different from that developed by Dalidovich and
 Phillips \cite {dal1, pi1} their approximations remain valid for
 our model because are related to the replica-trick method, and of
 the propagators of the replicated systems, which are the same for
 both models. On the other hand, the existence of the dissipation
 is essential for the occurrence of the metallic state as was
 first pointed out in \cite{wag}. However, if our model appear to
 be more tractable analytically by RG method, the main
 approximation of our model remain the dissipative form of the
 free propagator $G_0(\bf k,\omega_n)$.Such a form can be easy
 obtained if we adopt a model similar with this from \cite {ti}
 where the free propagator contains the scattering effects on the
 non-magnetic impurities, or the coupling to an Ohmic heat
 bath \cite{dal2}.
  This approximation is  well justified  by the
 results, and a RG-treatment containing more complicated form of the
 action is difficult to be controlled.
 The higher order
 corrections mentioned in  \cite{dal1} leads the terms
 proportionally to $u_0q^2$ which appear also in our
 calculations, but we cannot consider these corrections
 as having the same origin.
  In fact the vertex corrections presented later by the
 authors in Ref. \cite {dal1} leads to the same qualitative behavior for the $T=0$
 case, the difference being the power of $r_0$ from the
 denominator, a result similar with the calculations from \cite{ti},
 performed for the case of layered quasi-two dimensional
 superconductor. In fact the generalization of our model
 for this case is very simple problem and we do not expect
 qualitatively new results. The origin of the driving
 parameter $r_0$ is another problem which can be discussed because
 this transition is a $T=0$ transition, and in fact is a Quantum
 Phase Transition (QPT). In the next section we used the results from
 \cite{dal1,dal2} for the calculation of conductivity, without giving
 many technical aspects, and using the same
 approximations. However, we mention that the calculations  will be
 performed taking the conductivity to one -loop order per replica as in the
 Gaussian approximation. The new point is that we consider the
 influence of the fluctuations near the critical region  using
  $q(l^{\star})$ where $l^{\star}$ is given by  Eq.(17).

\section{Conductivity}

 The Kubo formula for a disordered system has the general
 form \cite{dal1}:
\begin{widetext}
\begin{equation}
\sigma(i\omega_n)=\frac{8e^2}{m \omega_n}T\sum_{a,b,\omega_l} \int
\frac{d^2{\bf k}}{(2\pi)^2}[G_{a,b}^{0}({\bf
k},\omega_l)\delta_{a,b}-2k_{x}^2G_{a,b}^0({\bf
k,\omega_l})G_{a,b}^0({\bf k,\omega_l+\omega_n})]
\end{equation}
\end{widetext}
Following the way from \cite{dal1} we transform this equation in
the simple form:

\begin{equation}
\sigma(\omega_n)=\frac{8q(l^{\star})e^2}{\omega_n}\int
\frac{d^2{\bf k }}{(2\pi)^2}k_x^{2}G_{0}^2({\bf k},0)[G_0({\bf k
},0)-G_0({\bf k},\omega_n)]
\end{equation}

In this point we mention that, differently from  \cite{dal1} we
considered the $l$-dependence of the effective order parameter $q$
for the disordered state in order to take into consideration the
influence of the interaction. This is not necessary in the case of
the Gaussian approximation mention above. Using now the simple
expression for $G_0$ and the Eq.(17) for $l^\star$ from Eq.(19)
and  $q(l^\star)=qe^{4l^\star}$  we get:
\begin {equation}
\sigma(\omega =0,T\longrightarrow 0 )=const\frac{q
\eta(1+u_0q)^2}{r_{0}^4}
\end{equation}

  This result shows that for $u_0=0$,which is Gaussian
  approximation we get the same result as in Ref \cite{dal1},
  and the metallic state is robust to the interaction between
  bosons.

\section{Acknowledgments}

 The work was partially supported by Romanian Ministry of
Education and Science. One of the authors (MC) is glad to thank
Professor Philip Phillips for stimulating discussions on this
problem.


\begin{thebibliography}{99}
\bibitem{exp1}H. M. Jager, D. B. Haviland, B. G. Orr, and A. M. Goldman, Phys. Rev. B{\bf 40}, 182 (1989).
\bibitem{exp2} D. Ephron, A. Yazdani, A. Kapitulnik and, M. R.
Beasley, Phys. Rev. Lett. {\bf 76}, 1529 (1996).
\bibitem{exp3} N. Manson and A. Kapitulnik, Phys. Rev. Lett {\bf 82},5341 (1999).
\bibitem{das} D. Das and S. Doniach,
Phys. Rev. B {\bf 60}, 1261  (1999).
\bibitem{fat} R. Fazio and G. Sch\"{o}n, Phys. Rev. B {\bf 43}, 5307 (1991)
\bibitem{wag}  K. Wagenblast,  A. van Ottero, G. Sch\"{o}n, and G. Zimanyi
Phys. Rev. Lett. {\bf 78} , 1779 (1997)
\bibitem{dal1} D. Dalidovich and P. Phillips, Phys. Rev. Lett.{\bf
89}, 27001 (2002).
\bibitem{pi1} P. Phillips and D. Dalidovich, Science {\bf 302},  243 (2003).
\bibitem{aa}  A. A. Abrikosov and S. I. Moukhin, J. Low Temp.  Phys.{\bf 33}, 207
(1978).
\bibitem{me} M. V. Medvedev, and A. V. Zaborov, Pys. Stat. Sol. (b){\bf
79}, 379  (1977).

\bibitem{bu1} G. Busiello, L. De Cesare, and I. Rabuffo, Phys. Rev. B {\bf 28}, 6463 (1983).

\bibitem{bu2}G. Busiello, L. De Cesare, and I. Rabuffo, Phys. Rev.B {\bf 29}, 4189 (1984).

\bibitem{da1} K. Damle and S. Sachdev, Phys. Rev. B {\bf 56}, 8717
(1997); Phys. Rev. B {\bf 57}, 8307  (1998).
\bibitem {sa} S. Sachdev, \textit{Quantum Phase Transitions  } (Cambribge University Press,
Cambridge, 1999).
\bibitem {ti}  I. Tifrea, D. Bodea, I. Grosu, and M. Crisan, Eur.
Phys. J. B {\bf 36}, 377  (2003).
\bibitem{fi} M. P. A. Fisher and G. Grinstein, Phys. Rev. Lett.{\bf
60}, 208  (1988).
\bibitem{ea} S. F. Edwards and P. W. Anderson, J. Phys. F {\bf 5}, 965
(1975).
\bibitem{dal2} D. Dalidovich and P. Phillips, Phys. Rev. B {\bf
63},  224503 (2001).

\end{thebibliography}
\end{document}